\documentstyle[multicol,epsfig,prb,aps]{revtex}

\ifpreprintsty
\def\multb{ }
\def\multe{ }
 \else
\def\multb{ \begin{multicols}{2}}
\def\multe{ \end{multicols}}
 \fi

\begin{document}
\draft
\preprint{HEP/123-qed}
\title{Ab-initio investigation of the covalent bond energies in the
  metallic covalent superconductor MgB$_2$ and in AlB$_2$}

\author{G. Bester and M. F\"ahnle}

\address{Max-Planck-Institut f\"ur Metallforschung, Heisenbergstr. 1, \\
D-70569 Stuttgart, Germany}

\date{\today}
\maketitle
\begin{abstract}
The contributions of the covalent bond energies of various atom pairs to the 
cohesive energy of MgB$_2$ and AlB$_2$ are analysed with a variant of
our recently
developed energy-partitioning scheme for the density-functional total
energy. The covalent bond energies are strongest for the intralayer B-B
pairs. In contrast to the general belief, there is also a considerable
covalent bonding between the layers, mediated by the metal atom.
The
bond energies between the various atom pairs are analysed in terms of
orbital- and energy-resolved contributions.
\end{abstract}


\multb 

\section{Introduction}

The ceramic compound magnesium diboride, MgB$_2$, has the highest
superconducting critical temperature T$_{\rm c}$ = 39 K ever reported for a binary
system. The structure is graphite-like, i.e., it consists of honeycomb layers
of B separated by triangular metal planes. Its discovery \cite{nagamatsu01a}
has triggered a tremendous activity to investigate the electronic
\cite{kortus01a,belashchenko01a,an01a,kurmaev01a,ravindran01a,medvedeva01a,kong01a,satta01a}
and phononic 
\cite{kortus01a,an01a,kong01a,satta01a,bohnen01a} 
properties as well as the electron--phonon
coupling 
\cite{kortus01a,an01a,kong01a,bohnen01a} for MgB$_2$ and for related binary
and ternary borides. The main electronic characteristics are: 
\begin{itemize}
\item The states at the Fermi level E$_{\rm F}$ are dominantly derived from
  boron $p$ orbitals. The B $p_{x,y}$ $\sigma$-bands are responsible for the 
  strong covalent bonding in the B layers and have mainly 2D character, i.e.,
  only very little $k_z$ dispersion, whereas the B $p_z$ $\pi$-bands have 3D
  character.
\item Electrons are transferred from the Mg orbitals to the B orbitals. As a
  result, there is a partial ionic character, but these electrons are not well 
  localized at the B sites, they are rather distributed over the whole crystal.
\item The Mg atoms are responsible for a downward shift of the B $\pi$-bands
  relative to the $\sigma$-bands, as compared to the corresponding C bands in
  graphite. As a result, there is a partial occupation of the $\pi$-bands and a
  hole doping of the $\sigma$-bands. Coupling of the $\sigma$-holes to the
  intralayer B bond-stretching phonon modes dominates the electron -phonon
  coupling and is responsible for the high T$_{\rm c}$ \cite{an01a,kong01a}.
\item The compound is hold together by dominant intralayer B-B covalent bonds,
  dominant interlayer metallic bonds and by a substantial ionic
  character. There are indications for an additional interlayer B-Mg
  contribution to the bonding, e.g., the slight increase in the $k_z$
  dispersion of the B $\sigma$-bands as compared to a hypothetical structure
  where the Mg atoms are removed. When going from MgB$_2$ to AlB$_2$, the
  valence charge-difference plots show a  stronger directional M-B bond
  \cite{ivanovskii00a} (M=metal atom), and our own calculations show a slightly
  stronger $k_z$ dispersion of the $\sigma$-bands. 

\end{itemize}

It is the objective of the present paper to analyse and compare the covalent
bonding properties of  MgB$_2$ and AlB$_2$ quantitatively within a variant 
of our recently developed energy-partitioning 
scheme \cite{bornsen99a,bornsen00a}. A first attempt in this direction was made
by Ravindran {\em et al.} \cite{ravindran01a} by means of the
crystal-orbital-Hamilton population \cite{dronskowski93a} (COHP). It will be
shown below, however, that for the use of non-orthogonal basis functions the
COHP do not have a well-defined physical meaning and cannot be used to compare
quantitatively the covalent bonding properties between various structures.

\section{The Energy-Partitioning Scheme}
\label{EnergyPartScheme}

Starting point of the discussion is the expression for the total energy
E$_{\rm tot}$ from the density functional theory in local density
approximation,
\begin{eqnarray}
\label{EtotDFT}
  E_{\rm tot} = \sum_n &f_n& \varepsilon_n - 
\int n({\bf r})v_{\rm eff}({\bf r}) {\rm d}^3r  \nonumber \\
 &+& E_{\rm H} + E_{\rm xc} + \int n({\bf r})v_{\rm ext}({\bf r}) {\rm d}^3r 
+ E_{\rm ii} \, ,
\end{eqnarray}
with the occupation numbers $f_n$ and the eigenvalues $\varepsilon_n$ of the
Kohn-Sham single-particle wave functions $\Psi_n$, 
the electron density $n({\bf r})$, the effective potential 
$v_{\rm eff}({\bf r})$, the Hartree energy $E_{\rm H}$, 
the exchange-correlation energy $E_{\rm xc}$, the potential 
$v_{\rm ext}({\bf r})$ of the nuclei (or of the ionic cores in the case of a
pseudopotential calculation) and the interaction energy 
$E_{\rm ii}$ between the nuclei (or between the ionic cores). 
The hope is that the trends for the total energies are already
well described by the first term (band-structure energy $E_{\rm band}$) when
comparing various systems. This assumption is made implicitly in the common
practice to discuss the energetics via the electronic density of states. An
energy-partitioning scheme is a tool to analyse $E_{\rm band}$ further. 

For systems with covalent bonding the crystal
wavefunctions $\Psi_n$ are best represented by a minimal set of well-localized
orbitals $\varphi_{i \alpha}$ (with character $\alpha$ described by the angular
and magnetic atom quantum numbers $l$ and $m$) attached to the various atoms 
$i$ (tight-binding representation),
\begin{equation}
\label{TBrep}
  \Psi_n = \sum_{i \alpha} c_{i \alpha}^n \varphi_{i \alpha} \qquad .
\end{equation}
The band-structure energy is calculated by inserting (\ref{TBrep}) into
\begin{equation}
  \sum_n f_n \varepsilon_n = \sum_n f_n \langle \Psi_n | \hat{\rm H} | \Psi_n 
\rangle
\end{equation}
with the Kohn-Sham Hamiltonian $\hat{\rm H}$. The remaining terms 
on the right-hand
side of eq. (\ref{EtotDFT}) which we abbreviated by $D$ in
Refs. \onlinecite{bornsen99a,bornsen00a} are calculated 
by approximating $n({\bf r})$ by
a superposition of atomic charge densities, yielding \cite{harris85a} 
\begin{eqnarray}
\label{HarrisDTerm}
  D = E_{\rm pair} &+& E_{\rm mb} + E^{\rm free\; atom} \nonumber \\ 
 &-& \sum_{i\alpha} 
N_{i\alpha}^{\rm free\; atom} H_{i\alpha i\alpha}^{\rm free\; atom} \qquad.
\end{eqnarray}
Here $E_{\rm pair}$ is a pair potential term,  $E_{\rm mb}$ a many-body
potential term which is small for nearly charge neutral systems, 
$E^{\rm free\; atom}$ denotes the total energy of the free atoms before being
condensed to the crystal, and the $N_{i\alpha}^{\rm free\; atom}$
are the occupation numbers for the energy levels 
$H_{i\alpha i\alpha}^{\rm free\; atom}$.
We now add to and subtract from $E_{\rm tot}$ the terms 
\[
\sum_n  \sum_{i\alpha, j\beta} \left[ c_{i\alpha}^n \left( c_{j\beta}^n
\right)^\ast O_{j\beta i\alpha} -  N_{i\alpha}^{\rm free\; atom} 
\delta_{j\beta, i\alpha}  \right] H_{i\alpha i\alpha}^{\rm free\; atom}
\]
and
\begin{equation}
  \sum_n \sum_{i\alpha, j\beta}c_{i\alpha}^n \left( c_{j\beta}^n
\right)^\ast O_{j\beta i\alpha} H_{i\alpha i\alpha} \qquad ,
\end{equation}
where $O_{j\beta i\alpha}$ and $H_{j\beta i\alpha}$ denote the elements of the 
overlap and Hamiltonian matrix. 

Rearranging the terms in an appropriate manner yields the cohesive energy
\begin{eqnarray}
\label{Sumofall}
  E_{\rm c} &=& E_{\rm tot} -  E^{\rm free\; atom} \\ \nonumber
&=& E_{\rm prom} + E_{\rm cf} 
+ E_{\rm polar} + E_{\rm cov} + E_{\rm pair} + E_{\rm mb} \quad .
\end{eqnarray}
The first term is the promotion energy
\begin{equation}
  E_{\rm prom} =  \sum_{i \alpha} \left( q_{i \alpha} - N_{i \alpha}^{\rm free\;
    atom} \right) H_{i\alpha i\alpha}^{\rm free\; atom}
\end{equation}
with Mulliken's gross charge \cite{mulliken55a}
\begin{equation}
  q_{i\alpha} =  \sum_{j\beta} \sum_n f_n \, c_{i\alpha}^n \left( c_{j\beta}^n
\right)^\ast O_{j\beta i\alpha} \qquad .
\end{equation}

This term describes the cost in energy when starting the condensation process
from free atoms and then redistributing the electrons among the various
orbitals from the occupation numbers $N_{i\alpha}^{\rm free\; atom}$ to the
occupation number $q_{i\alpha}$ found in the crystal and characterized by
$q_{i\alpha}$\cite{TBBbondcomment}. The second term is the crystal-field term
\begin{equation}
  E_{\rm cf} = \sum_{i\alpha} q_{i\alpha} \left( H_{i\alpha i\alpha} -
  H_{i\alpha i\alpha}^{\rm free\; atom} \right) \qquad ,
\end{equation}
which describes the change in energy due to a shift of the on-site energies
when the atoms are condensed to form the crystal so that the potential acting
on an electron at atom $i$ is not just the atomic potential of this atom but
the environment-dependent crystal potential. The polarization energy 
\begin{equation}
  E_{\rm polar} = \sum_{n,i,\alpha,\beta} f_n \, c_{i\alpha}^n \left( c_{i\beta}^n
\right)^\ast \left[  H_{i\alpha i\beta} - \delta_{i \alpha}^{i\beta}
H_{i\alpha i\alpha} \right] 
\end{equation}
describes the change in energy due to the hybridization of orbitals localized
at one atom when the atom is embedded in the crystal. Finally, the covalent
bond energy $E_{\rm cov}$ is the change in energy arising
from the hybridization of orbitals localized at different atoms, 
\begin{equation}
  E_{\rm cov} =  \sum_{i\alpha, j\beta \atop j\ne i} E_{{\rm cov}, i\alpha j\beta}
\end{equation}
with
\begin{eqnarray}
  E_{{\rm cov}, i\alpha j\beta} &=& \sum_n f_n \, c_{i\alpha}^n \left(
  c_{j\beta}^n \right)^\ast \left[  H_{j\beta i\alpha} -  O_{j\beta i\alpha}
\, \overline{\varepsilon_{j\beta i\alpha}}  \right]  \\
\overline{\varepsilon_{j\beta i\alpha}} &=& \frac{1}{2} \left(
H_{i\alpha i\alpha} + H_{j\beta j\beta} \right) \qquad . 
\end{eqnarray}
$E_{{\rm cov}, i\alpha j\beta}$ can be subdivided further into energy-resolved 
contributions,
\begin{eqnarray}
   E_{{\rm cov}, i\alpha j\beta}(E) = \sum_n \delta (E-\varepsilon_n)\, f_n \, 
c_{i\alpha}^n  \left( c_{j\beta}^n \right)^\ast \nonumber \\ 
\times \left[  H_{j\beta i\alpha} -  O_{j\beta i\alpha} \,
\overline{\varepsilon_{j\beta i\alpha}}  \right] \qquad .
\end{eqnarray}
$ E_{{\rm cov}, i\alpha j\beta}(E)$ is negative (positive) for bonding
(antibonding) states. The respective quantity integrated up to a certain
energy $E$ will be referred to as 
\begin{equation}
  IE_{{\rm cov},i\alpha j\beta} (E) = \int_{-\infty}^E  E_{{\rm cov}, i\alpha
    j\beta}(E')\, {\rm d}E' \quad .
\end{equation}

The energy-partitioning scheme discussed above has the following very
attractive property: In a band-structure calculation which deals with an
infinitely extended periodic system the average effective potential does not
have a physical meaning, and it is therefore set to an arbitrary value which
is the same for different crystal structures. In order to be physically
meaningful the total energy $E_{\rm tot}$ and the considered terms of an 
energy-partitioning scheme for a
band-structure calculation therefore must be invariant against a constant
shift of the effective potential of the band-structure
calculation. 
This is fulfilled for $E_{\rm tot}$ (and hence also for $E_{\rm c}$) which
becomes obvious from eq.(\ref{EtotDFT}): Shifting $v_{\rm eff}$ by $\Phi_0$
yields opposite shifts for the first two terms which therefore compensate each 
other (the remaining terms can be calculated without ambiguity, see for
instance, Ref. \onlinecite{ihm79a}.
Furthermore the terms $E_{\rm prom}$, 
$E_{\rm polar}$ and $E_{\rm cov}$ of eq. (\ref{Sumofall}) as 
well as their atom- and orbital-resolved contributions (and in addition the
respective energy-resolved contributions to $E_{\rm cov}$) are all invariant.
For instance, if the potential is shifted by a constant $\Phi_0$, then the
matrix elements $H_{j\beta i\alpha}$ are transformed into  $H_{j\beta i\alpha}
+  \Phi_0 O_{j\beta i\alpha}$ and $H_{i\alpha i\alpha}$, 
$\overline{\varepsilon_{j\beta i\alpha}}$ into $H_{i\alpha i\alpha} +
\Phi_0$, $\overline{\varepsilon_{j\beta i\alpha}} + \Phi_0$ because  
$O_{i\alpha i\alpha} = 1$, so that $\Phi_0$ drops out of the covalent bond
energy.  
Because $E_{\rm c}$ is also invariant, this must hold for the sum of the 
terms $E_{\rm pair}+E_{\rm mb}+E_{\rm cf}$, too. 
However, we cannot calculate separately, for instance, the crystal-field
term $E_{\rm cf}$ in a band-structure calculation, because the matrix 
element $H_{i\alpha i\alpha}$ is 
shifted by the shift of the effective potential of the crystal whereas
$H_{i\alpha i\alpha}^{\rm free atom}$ 
is not (because for the calculation of the latter quantity the
effective potential is always normalized to zero for distances far from the
atom). It is therefore physically meaningful only to discuss the terms 
$E_{\rm prom}$, $E_{\rm polar}$, $E_{\rm cov}$  and $E_{\rm pair}+E_{\rm
  mb}+E_{\rm cf}$. The covalent bond energy thereby is the
only term which involves matrix elements for orbitals on different atoms, and
therefore it clearly represents the contribution of the interatomic bonding to 
the band-structure energy. In the following we will confine ourselves to the
discussion of this term.

It should be noted that in our former version of the energy-partitioning
scheme \cite{bornsen99a,bornsen00a} the various terms have been arranged in a
slightly different manner, arriving \cite{FormerDefComment}  at the 
equivalent expression
\begin{equation}
    E_{\rm c} =  \widetilde{E}_{\rm prom} +  
 \widetilde{E}_{\rm cf}  +  \widetilde{E}_{\rm cov} + E_{\rm pair} +
    E_{\rm mb} \qquad ,
\end{equation}
with
\begin{equation}
\widetilde{E}_{\rm prom} =  \sum_{i \alpha} \left( q_{i \alpha} - N_{i
  \alpha}^{\rm free\; atom} \right) H_{i\alpha i\alpha} \qquad , 
\end{equation}
i.e., $H_{i\alpha i\alpha}^{\rm free\; atom}$ of $E_{\rm prom}$ has 
been replaced by  $H_{i\alpha i\alpha}$,
\begin{equation}
\widetilde{E}_{\rm cf} = \sum_{i\alpha} N_{i\alpha}^{\rm free\; atom} 
\left( H_{i\alpha i\alpha} - H_{i\alpha i\alpha}^{\rm free\; atom} \right) 
\qquad ,
\end{equation}
i.e., the $q_{i\alpha}$ of $E_{\rm cf}$ have been replaced by  
$N_{i\alpha}^{\rm free\; atom}$, and
\begin{equation}
\widetilde{E}_{\rm cov} =  \sum_{i\alpha, j\beta} E_{{\rm cov}, i\alpha
  j\beta} \qquad ,
\end{equation}
i.e., the on-site contributions $i\alpha i\beta$ have not been excluded from
the covalent bond energy. When comparing the definitions $E_{\rm prom}$, 
$E_{\rm cf}$, with the definitions  $\widetilde{E}_{\rm prom}$, 
$\widetilde{E}_{\rm cf}$, it becomes obvious that they correspond to a 
different 
succession of processes in a gedanken experiment for the condensation of free
atoms to the crystal. In the first case we promote the electrons by
redistributing them among the various orbitals of the free atoms and then we
bring the free atoms in the crystal positions (without allowing for a
redistribution of the charge densities) and experience a change in energy
described by $E_{\rm cf}$ due to a shift of the on-site energies in the crystal
potential. In the second case we freeze the occupation numbers
$N_{i\alpha}^{\rm free\; atom}$  when bringing the free atoms to the 
crystal positions and calculate the crystal-field shift $\widetilde{E}_{\rm
  cf}$ for these circumstances, and then we allow for a redistribution of the
electrons among the on-site energy levels in the crystal potential. We think
that the first case is closer to the commonly used definitions of the
promotion and the crystal-field energy, and we therefore prefer the new
variant of the energy partitioning scheme. Finally, we think that it is
reasonable to exclude the on-site hybridization contributions 
$E_{\rm cov, i\alpha i\beta}$ from the covalent bond energy because they do not 
describe interatomic interactions. In the new variant these terms enter the
polarization energy which also has a well-defined physical meaning.

It should be noted that $\widetilde{E}_{\rm prom}$, 
$\widetilde{E}_{\rm cov}$ and 
$\widetilde{E}_{\rm cf} + E_{\rm pair} + E_{\rm mb}$ are also 
invariant against a constant shift 
of the effective potential of a band-structure calculation, as well as 
the atom-, orbital-  and energy-resolved contributions of 
$E_{\rm cov}$. Again, $\widetilde{E}_{\rm cf}$ alone is not 
invariant and cannot be
calculated separately in a band-structure calculation.

In former papers the crystal-orbital-Hamilton population 
\cite{dronskowski93a},
\begin{equation}
COHP_{i\alpha j\beta}(E) = \sum_n \delta (E-\varepsilon_n)f_n 
c_{i\alpha}^n \left( c_{j\beta}^n \right)^\ast H_{j\beta i\alpha}\;,
\end{equation}
has been used to characterize the bonding properties. If orthonormal basis
functions are used, this quantity is identical to 
$E_{\rm cov, i\alpha j\beta} (E)$. However, in the chemical analysis very
often non-orthogonal basis sets are used. Then $COHP_{i\alpha j\beta}(E)$ is
not invariant against a constant shift of the effective potential and
therefore does not have a well-defined physical meaning in the context of a
band-structure calculation.

For systems with metallic bonding
the $\Psi_n$ are better represented by a set of plane waves rather than by
atom-localized functions.
Alternatively, the $\Psi_n$ again can be represented by a set of 
atom-localized functions 
also in this case, but then often orbitals have to be included which are not
occupied in the free atom in order to make the basis set more complete. Then
formally a covalent bond energy $E_{\rm cov}$ can be calculated even for a
nearly-free-electron system.
Whereas the term metallic conductivity is well defined, it is indeed a 
problem to discriminate between metallic and covalent bonding. As a working
hypothesis we define a covalent bonding as a bonding which is dominated by
the hybridization of those orbitals on various atoms which are 
already occupied in
the respective free atoms. Note that this does not necessarily mean that the
corresponding charge-density difference plot exhibits directionality, e.g., it
can be imagined that the $p_z$-orbitals do not necessarily create a 
charge density with
considerable directionality. Because our definition of $E_{\rm cov}$
 is a generalization of the covalent bond energy introduced by Sutton {\em et
   al.} \cite{TBBbondcomment} to the case of non-orthogonal basis sets we keep this 
 historical founded nomenclature ``covalent bond energy'' although this
 quantity may also contain metallic bonding contributions in the above defined
 sense. 

\section{Details of the calculations}

The calculations were performed with the ab-initio pseudopotential method
\cite{vanderbilt85a,meyer_prog}. Band structure and charge density difference
plots were practically identical to those of previous calculations
\cite{kortus01a,belashchenko01a,an01a,kurmaev01a,ravindran01a,medvedeva01a}. 
For the energy-partitioning analysis 
the $\Psi_n$ were projected \cite{sanchez-portal95a96a,kostelmeier99a,meyer_phd,bester01a}
onto a set of overlapping
atom-localized non-orthogonal orbitals.
For these orbitals we chose 
\begin{equation}
\label{DefderAtomarenBasisfkt}
\varphi_{i \alpha}({\bf r}) = 
f_{i l}(r)\, i^l\, {\rm K}_{lm}(\widehat{\bf r}) \qquad ,
\end{equation}
\begin{equation}
\label{ao_abschneiden}
f_{i l}(r) = C_{i l}\,
\phi_{i l}^{\rm PS}(\lambda_{i l}r) \left\{
\begin{array}{ll}
\big(1 - {\rm e}^{-\textstyle \gamma_{i l}(r_{i l}^{\rm cut}-r)^2} \big)
         & \mbox{for } r \le r_{i l}^{\rm cut} \\
0        & \mbox{for } r \ge r_{i l}^{\rm cut}, \\
\end{array}\right.
\end{equation}
where $C_{i l}$ is a normalization constant, $\phi_{i l}^{\rm PS}$ is the
radial pseudo-atomic wavefunction constructed according to Vanderbilt
\cite{vanderbilt85a}, $\lambda_{i l}$ denotes a contraction factor and 
$r_{i l}^{\rm cut}$ represents a cut-off length. The parameters  $\lambda_{i
  l}$, $\gamma_{i l}$ and $r_{i l}^{\rm cut}$ were selected in such a way
that the spillage 
\cite{sanchez-portal95a96a,kostelmeier99a,meyer_phd,bester01a}  was minimized, 
where the spillage characterizes the loss of the norm of the wavefunctions due 
to the incompleteness of the pseudo-atomic-orbital projection.
We confined ourselves to a
minimal set of s,p and d orbitals for Mg and B. The band-structure calculated 
with the projected wavefunctions was nearly identical to the original band
structure from the pseudopotential calculation for energies below and not 
too far above the Fermi level.

\section{Results and Discussion}

Table \ref{table1} represents the covalent bond energies for various
atom pairs in MgB$_2$ and  AlB$_2$. It should be recalled that the covalent 
bonding properties of the two materials may be compared only by the measure
$E_{\rm cov}$ and not by COHP because the latter quantity is not invariant
against a constant shift of the potential when nonorthogonal basis functions
are used, which is the usual case for a chemical analysis in terms 
of atom-like functions. The total covalent bond energy is larger in 
absolute value for
AlB$_2$ than for  MgB$_2$. As expected, the B-B intralayer covalent bond
energy is largest. Interestingly enough, it is larger in absolute value for
the Mg compound although the B-B distance is larger. Contrary to the general
belief, there is also a considerable covalent bond energy between the
layers (the B-M energy is only a factor of about 2-2.5 smaller) and even in
the M layers (the intralayer M-M energy is a factor of about 
two smaller than the
B-M energy). The B-M covalent bond energy increases when going from 
M=Mg to M=Al, and this is consistent with the smaller
c/a ratio of the Al compound. The B-M distance is in fact smaller for M=Al
than for M=Mg. The interlayer B-B (M-M) bonds are two orders of magnitude
(one order of magnitude) smaller than the respective intralayer bonds,
i.e., their contributions to the cohesion between the layers is much smaller
than the B-M contribution. Our calculations show that the covalent 
bond energies between all
further-distant atoms are considerably smaller than the nearest-neighbor
intralayer bonds and the nearest-neighbor B-M bonds. This means that a
nearest-neighbor bond model suffices to describe the two materials. This is
not at all trivial, because in intermetallic compounds, e.g., FeAl, CoAl and
NiAl the further distant bonds are essential \cite{bornsen00a}.

In Table \ref{table2} the covalent bond energies for the most important
atom pairs are further analysed by considering the dominant angular-resolved
contributions. 
In all cases, the p-p contributions are strongest and the s-s
contributions weakest. 
It becomes obvious from Table \ref{table2} that the $d$-orbitals make a
non-negligible contribution to the bonding between the B and the M
layer. Because the $d$-orbitals are not occupied in the respective free atoms
we would denote this as a ``metallic contribution'' according to our working
hypothesis of section \ref{EnergyPartScheme}. This metallic contribution is
stronger for AlB$_2$ than for  MgB$_2$. Nevertheless, the ``covalent
contribution'' is dominant also for the bonding between the layers, mediated
by the metal atom.

\begin{figure}
\centerline{
\epsfig{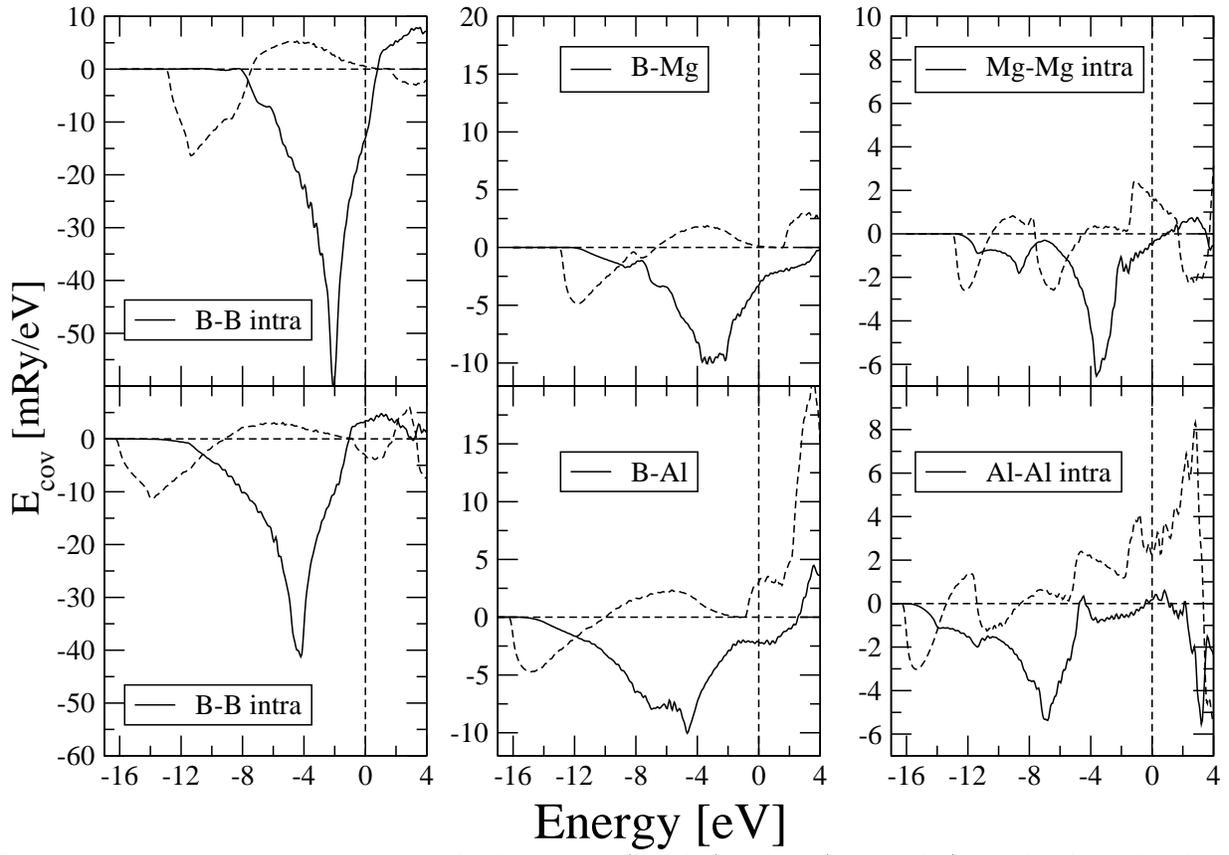}
}
\setlength{\columnwidth}{\linewidth} \nopagebreak
\caption{Energy resolved covalent bond energies for the p-p (full line) and 
  s-s (dashed line) contributions of various atom pairs in MgB$_2$ (upper
  panels) and AlB$_2$ (lower panels).}
\label{figure1}
\end{figure}

The energy-resolved covalent bond energies for the
p-p and s-s contributions of the B-B intra, B-M and M-M intra atom pairs are
shown in Fig. \ref{figure1}, 
for MgB$_2$ and AlB$_2$. 
The benefit of the energy-resolved  representation is that we can clearly
discriminate between bonding and antibonding states.
The s-s bonds are weakened
because both bonding and antibonding states are occupied.
Without the energy-resolved analysis we could erroneously assume that
the s-s covalent bond energy is low because the corresponding matrix elements
are small. For the other orbital pairs of Table \ref{table2} dominantly
bonding states are occupied, but there are slight differences between  
MgB$_2$ and AlB$_2$. For instance, the stronger B-B intra bond energy of
MgB$_2$ (Table \ref{table1}) results mainly from a stronger $p-p$ contribution 
(Table \ref{table2}), and the reason for this is that for the trivalent Al
part of the antibonding p-p states are occupied
(Fig. \ref{figure1}). Furthermore, the stronger B-M bond for M=Al (Table
\ref{table1}) again results mainly from a stronger $p-p$ contribution, and the 
reason for this is that the bonding p-p states are more filled for Al than
for Mg. 

\section{Conclusion}

We have represented a new variant of an energy-partitioning scheme for the
density-functional total energy which allows to define a covalent bond energy
which is invariant against a constant potential shift. This property is a
precondition for the comparison of the bonding properties in different 
systems 
within the framework of band-structure calculations. 

The atom-, orbital- and energy resolved contributions to the covalent bond
energy have been calculated for MgB$_2$ and AlB$_2$. One central result of the
calculations is that a nearest-neighbor bond model is sufficient to describe
the bonding properties of these materials. A working hypothesis has been
introduced to discriminate between a covalent and a metallic bonding
character. It has been shown that both in MgB$_2$ and AlB$_2$ there 
is a metallic
contribution to the bonding between the layers which is stronger for Al than
for Mg. The benefits of the energy-resolved representation are demonstrated
by discussing the differences between the trivalent Al and the divalent Mg. 

\section*{Acknowledgment}

The authors are indebted to R. Drautz for helpful discussion.

%
%

%
%

\begin{table}
\begin{tabular}{c||c|c|c|c|c|c}
        & B-B   & B-B   & M-M   & M-M   &  B-M & total \\
        & intra & inter & intra & inter &      &       \\ \hline
MgB$_2$ & -193.5& 1.3   & -37.6 & -2.7  & -73.7& -3031 \\
AlB$_2$ & -180.3 & 3.7 & -41.7 & 10.2 & -86.9 & -3179 \\
\end{tabular}
\caption{\label{table1} Covalent bond energies (in mRy) for various atom
  pairs.}
\end{table}  

\begin{table}
\begin{tabular}{c|ccc|cccc|c}
        &  \multicolumn{3}{c|}{B-B intra} & \multicolumn{4}{c|}{B-M}   & M-M
        intra \\ 
        & p-p & p-s & s-s & p-p & p-d & p-s & s-s & p-p \\ \hline
MgB$_2$ & -84.9 & -42.8 & -14.4 & -24.9 & -13.2 & -12.8 & -3.6 & -10.0 \\
AlB$_2$ & -72.9 & -39.5 & -14.2 & -30.8 & -17.6 & -14.1 & -1.7 & -12.3 
\end{tabular}
\caption{\label{table2} The dominant orbital-resolved bond energies (in mRy)
  for various atom pairs.}
\end{table}  

%
%

\multe

\begin{references}
\bibitem{FormerDefComment} This form is obtained from the energy-partitioning
  scheme of Refs. \onlinecite{bornsen99a,bornsen00a} when taking into account
  eq. \ref{HarrisDTerm}.
\end{references}
\end{document}